\documentclass[seceq,wrapft]{ptptex}
\notypesetlogo

\usepackage{graphicx}

\title{
Ultra-Luminous Sources in Nearby Galaxies }
\author{
Richard {\sc Mushotzky }}
\inst{NASA Goddard Space Flight Center, Greenbelt, MD 20771, USA}

\abst{
I briefly review much of the X-ray and optical data on the nature of the ULXs
in nearby galaxies. I present new results on radio emission, finding that the
radio is usually rather luminous and extended. I review the X-ray data on
timing and spectra. There is no direct evidence in the X-ray data for either
geometric or relativistic beaming and in 4 objects direct evidence against
beaming. I argue that the X-ray timing and spectral properties of these
objects, are in general, not good analogs of AGN or galactic black holes and
that the ULX may represent a new mode of accretion only rarely seen in other
objects.}

\begin{document}

\maketitle

\section{Introduction}

Since their discovery in observations of nearby galaxies with the Einstein
observatory (Fabbiano et al. 1989), the existence of a class of
ultra-luminous X-ray sources has been well documented. For the purposes of
this paper, we shall call a source ``ultra-luminous'' or  a ULX, if it is not
at the galaxy nucleus, it is spatially unresolved with Chandra or is  time
variable and if its observed bolometric luminosity exceeds the Eddington
limit for a 20$M_\odot$ black hole (2.8$\times 10^{39}$ ergs/sec). Because black holes of
greater than  20$M_\odot$ are not expected from ``normal'' stellar evolution (even
from very massive stars, Fryer and Kalogera 2001) and are clearly different
from the ``normal'' population of black hole X-ray sources as seen in the
Milky Way we believe that this is the appropriate luminosity to
consider. These objects are moderately numerous since they are in  $\sim$1/4 of
all galaxies (Colbert and Ptak 2002). They are of great interest since they
represent a population of {\it possible} intermediate mass (20--5,000$M_\odot$) black
holes (IMBHs) or a stage in the evolution of binaries not seen in the Milky
Way. The existence of such a population poses strong theoretical challenges
to their creation and fueling (talk by King this workshop), while if they are
not IMBHs their high total luminosity and unique spectral and timing
properties pose strong challenges to physical models for their radiation
mechanism (Ebisawa et al., these proceedings, Kubota et al., these proceedings).  In defining the
luminosity of these objects one must be careful to calculate the bolometric
luminosity. For many of the  models fit to the ULX X-ray spectra the 0.5--10
keV observed band contains less than
40\% of the total luminosity. Thus many of these objects are more luminous than
reported in much of the literature.

As we will show in this paper many of the ULXs have properties not shared by
AGN or galactic black holes and that, as a class, they show a wide variety of
spectral and temporal behaviors which suggest that they may be several types of
object and not represent a uniform population.

\section{Classes of models}

After the discovery of these objects many models were suggested to explain
their
origin. They basically fall into 3 classes: supernova remnants,
non-isotropic emission from ``normal'' black hole binaries and ``normal''
emission from intermediate mass black holes.

\subsection{Supernovae in dense  environments}
 These objects are known to exist and can
have observed luminosities of   $L_x\sim10^{38}$--$2\times10^{40}$ ergs/sec (e.g. SN 1995N, Immler
and Lewin 2002). Their high luminosity is due to the rapid dissipation of the
supernova shock energy in a dense environment and thus these objects are
associated with historical supernova. Their luminosities are observed to
decline monotonically over a few years (Immler and Lewin 2002).  However, there
exist anomalous  SNR, such as the SNR in NGC4449, whose XMM X-ray spectrum are
clearly thermal and  are associated with  optical objects which have classical SNR
emissicon line ratios, and which have very high luminosities, but are not
associated with historical remnants and whose luminosity does not decline much
with time (Patnaude and Fesen 2003).

The other two possibilities  are ``theoretical'' objects postulated to exist to
produce the high observed luminosities of the ULX.

\subsection{Non-isotropic emission}
 The high luminosities of the ULX may be explained by non-isotropic emission
from X-ray binaries: either from a ``normal'' high-mass X-ray binaries (HMXB) or
micro-blazars (beamed emission from relativistic jets). In both of these
classes of  models the observed high luminosity is due to either geometrical or
relativistic beaming (K\"{o}rding et al.), a short lived super-Eddington phase
(King 2003) or a stable super-Eddington luminosity due to a geometrical
arrangement of emission and absorption regions which allows the photons to
escape without interacting with surrounding material (Begelman 2002). While
micro-quasars certainly exist in the Milky Way (see Mirabel, these proceedings) the ULXs can be 1000 times more luminous and would be a rather
different type of object.

\subsection{Accretion onto massive objects}
The third class of models invokes the existence of very massive objects in
order to reproduce the bolometric luminosity of these objects with ``normal''
accretion processes. To not exceed the Eddington limit  one needs a mass
between 20--2000$M_\odot$ and thus this class of models posits the existence of
intermediate-mass black holes (IMBHs). There also  exist ``lost'' LLAGN which
are low-luminosity AGN which are not detected by optical techniques and are not
in the ``obvious'' nucleus of the galaxy (e.g. NGC3256 Neff et al. 2003). These
objects are presumed to have been displaced from the apparent nucleus via a
strong merger. Their relative lack of optical emission lines is not unusual
amongst Chandra selected AGN (Mushotzky 2004).

\section{What is the available data?}
There is now a considerable body of data  from archival Rosat (Colbert and Ptak
2002), Chandra and XMM data (Humphrey et al., Swartz et al., Ptak et al.)
about the relative  numbers, X-ray luminosity, luminosity function and X-ray
colors of the ULXs in a reasonable number of nearby galaxies. There have been
extensive searches for counterparts in other wavelength bands (optical, 
near-IR and radio).  A reasonable number of objects have moderate to high quality
X-ray CCD spectra from ASCA, Chandra and XMM in the 0.3--10 keV band (Kubota 
et al., these proceedings, Ebisawa et al., these proceedings).

There is a large amount of data based on X-ray time variability on long (years) to short
(seconds) time scales (e.g. Roberts and Warwick 2000) for the brighter sources,
but with very uneven time sampling.  There are extensive studies of the
correlations of ULX properties with galaxy properties (Colbert et al. 2003,
Gilfanov et al. 2003, Gilfanov, these proceedings, Swartz et al. 2003).

\section{The critical issues}
The main issue in this field is, clearly, ``what is the true nature of the ULX
sources and what is their emission mechanism?'' This boils down to whether
$ M>20M_\odot$
objects really exist, and if they do not, how can one produce the observed very
high luminosities.

\subsection{Arguments against $M>20M_\odot$   objects} 
As discussed in detail by King and co-workers (talk by King this workshop) there
are several substantial  arguments against $M>20M_\odot$   objects being the origin of
most of the ULXs. These arguments
 can be grouped into two parts.

\subsubsection {Astronomical}

There are severe  difficulties in forming such massive objects and providing
sufficient fuel to feed them. Considering their apparent high accretion rate
($10^{-8}$--$10^{-6} M_\odot$/yr) the lifetime of any companion must be short and thus, there
must be many ``quiescent''  ULX for each luminous one.

If these objects have  $M>20 M_\odot$ they cannot form from stellar evolution of
single ``normal'' massive stars and their origin must lie in collective
phenomena (Miller and Colbert  2003).
 It is difficult to
form such objects in classical binary stellar evolutionary scenarios unless the
companion (which provides the ``fuel'') is also massive and thus has a short
lifetime. If the objects are not formed as binaries, they must ``acquire'' a
stellar companion via capture, which  maybe difficult.  If their true
luminosities are so high one has to consider  possible ablation of a companion
which should further shorten the life of the system.

\subsubsection{Statistical}

ULX tend to be associated with recent star formation regions and their total
numbers are correlated with the star formation rate (Gilfanov et al. 2003). A
small number have possible optical associations with luminous stars, indicating
that they are associated with massive, short lived stellar objects. Some show
transitions similar to that seen in galactic black holes indicating that they
are similar objects. As discussed in detail by Gilfanov (these proceedings) the
overall luminosity function of galaxies does not have a ``feature'' associated
with the ULXs indicating that there is smooth transition between objects more
and less luminous that a $20M_\odot$ object radiating at the Eddington limit.

However in all models in which the ULXs are less than $20M_\odot$ the observed
luminosity must be due either to beaming in an intrinsically Eddington limited
object, or via radiation from a super-Eddington object by some means (see
Begelman 2002, King 2003 for such possibilities).

\subsection{Arguments against $M<20 M_\odot$ objects}
  While such astronomical/theoretical arguments seem daunting there are many
direct observations which seem to rule out beaming or super-Eddington
luminosities in these objects.

What are the arguments against ULX being ``normal'' $M<20 M_\odot$ objects?

\subsubsection{Data}

The  X-ray spectra of the ULX are often not like AGN or normal galactic black
holes (see below)  indicating that they are not simple analogs of these
objects.  The ULX can have state transitions like ``normal'' black holes
(Kubota et al.  2001)  or  in the opposite sense from the vast majority of
galactic objects (Fabbiano et al. 2003, Dewangan et al. 2004). Their bolometric
luminosities can reach 1000 $L_{\rm edd}$ for a 1 solar mass object making 
beaming or
super-Eddington models rather extreme.

There is direct evidence against beaming  in several objects: viz the QPO in
M82, the  broad Fe lines in two ULX M82 (Strohmayer and Mushotzky 2003)  and
the Circinus galaxy  (Weisskopf et al. 2003), and the evidence for eclipses in 2
objects (the Circinus galaxy ULX  and one of the ULX in M51 
(Liu et al. 2002)).
At least one object has a break in the power density spectrum (PDS) at the
frequency predicted for $M\sim 1000 M_\odot$ objects (Cropper et al. 2004).

As shown later in this paper many of the ULXs have associated, very luminous
extended radio sources, very different from any associated with Milky Way black
holes. There are a few ULXs with highly luminous photoionized nebulae around
them (Pakull and Mironi 2002). If these nebulae are photoionized by the ULX, as
indicated by their association and unusual optical spectra,  one requires a
high total  ULX luminosity.  There is also a general lack of stellar optical
identifications; in many of the galaxies observed with both HST and Chandra the
sensitivity threshold of HST is such that   luminous O star counterparts would
have been detected.  While the ULXs are associated with star forming regions,
they lie near but not in them.  This requires,  that if the ULX are created in
these star forming regions, that the ULXs have a high space motion (Kaaret et
al. 2004) but ``carry around'' the extended radio sources and optical nebulae,
which to this author seems unlikely. Quite a few of the ULXS have ``soft''
components well fit by a low $kT$ black body whose temperature and flux are
consistent with a high mass (Miller, these proceedings).

\subsubsection {Theory}

From a theoretical point of view there are no known mechanisms for  the
required beaming ($>100$) in the most luminous objects other than relativistic
effects. The observed X-ray spectra and time variability behavior does not
resemble that of known relativistically beamed objects (such as Bl Lacs).  In
many sources (e.g. NGC2276 Davis and Mushotzky 2004) the observed luminosity
function lacks the large number of lower luminosity objects (K\"{o}rding, these
proceedings) required in a relativistic beaming scenario  and   thus is not
consistent with beaming. The ULX are  not ``ultraluminous'' in other wavelength
bands like AGN are, thus placing strong constraints on models (King 2003) in
which the actual luminosity is super-Eddington.  While quasars and Seyfert
galaxies radiating near the Eddington limit (Czerny et al. 2003) appear to have
a ``universal'' broad band spectrum that is not seen in the ULXs, reducing the
likelihood that these objects are radiating near the Eddington limit.

\section{Where do they come from?}

If the ULX are really $M\sim100M_\odot$ objects, there are, at present, two possibilities
for their origin. The ULX may be created in the early universe;  detailed
calculations of the first stars to form (e.g. Abel 2003) indicate that
$M\sim 200$--$1000M_\odot$ objects should be created, and that they are numerous and lie in
regions that will later become galaxies.

Alternatively, the  ULXs could be created in dense stellar regions (e.g.
globular clusters Miller and Colbert 2003 or dense star clusters (McMillan and
Portegies Zwart 2003)). However since most ULXs today are not observed to be in
dense clusters this requires that either the ULXs are ejected or that the
clusters dissolve or otherwise become invisible. If they are not supermassive
objects then they must represent an unusual stage of stellar evolution and/or
some new mode of accretion.

\section{How do we determine their nature?}
If  ULXs are intermediate mass  black holes they should have properties that
scale from the better studied  AGN at high mass and galactic black holes at low
mass. These properties include the detailed nature of their time variability,
the spectral form of their broad band X-ray spectra and the detailed spectra in
X-ray/radio/optical bands. In particular, one would like to measure properties
that scale with mass such as the temperature of the low energy component, where
one expects that $kT \sim M ^{1/4}$, and the characteristic time scale of 
X-ray
variability (Markowitz et al. 2003) where the break in the power density
function scales with mass. As we show in this paper some of these scalings
have been observed for some of the objects but most of the ULXs do not show the
expected relationships.

\section{What can we learn from optical associations?}
Much progress in understanding the nature of both galactic and extragalactic
X-ray sources has come from detailed optical ``follow-up'' observations and the
original hope was that similar programs, that succeeded so well in identifying
the nature of galactic and extragalactic X-ray sources, would be the key to
understanding the ULXs. Unfortunately this has turned out to be considerably
harder than first hoped. If unique ``identification'' of optical counterparts
is obtained one could  estimate the  masses of ULXs, their evolutionary
history and discriminate between models.  If  optical emission line nebulae are
associated with the  ULX they can be used as calorimeters to derive true
isotropic luminosities of the objects, constrain the lifetime of the ULXs, the
form of the ionizing continuum and place them in their evolutionary context.

\subsection{Optical  nebulae}

Most of the reported optical nebulae which are associated with the ULXs (Pakull
and Mironi 2003) are very big $\sim200$--600 pc, and very energetic, with 
kinetic
energies $\sim10^{52}$--$10^{53}$ ergs/sec  much more than standard SNR. There are at least
2 variable ULXs associated with nebulae that were thought to be supernova
remnants, based on the nature of the optical emission lines (Roberts et al.
2003a and b). The detailed optical spectra of these nebulae can, in
principle, distinguish shock vs photoionization origins for the optical
emission lines and determine whether they may be excited by the ``central''
X-ray source. Many of the nebulae are rather  unusual,  having both strong
[OIII], NeIII and HeII emission lines as well as lower ionization lines of
[OI] and [SII]. The most interesting case so far is the nebulae associated with
the ULX in the dwarf galaxy Holmberg II (Pakull and Mironi 2003) which shows
strong HeII 4686 and [OI] 6300. Since HeII is produced by recombination it
requires a strong high energy source of photoionization with an isotropic
luminosity of greater than 3$\times10^{39}$ ergs/sec to produce the observed optical
emission line spectrum. 

Over 4 ULX are inside  bubble-like nebulae, 200--400 pc
in size, much larger than SNR in the Milky   Way. It is not yet clear if the
nebulae associated with the ULXs are highly unusual and/or how statistically
likely a chance coincidence between such a large and luminous nebula with a
random ULX would be.

\subsection{Optical stars}

Even with the precision Chandra positions there is often no unique optical
counterpart to the ULX (Cropper et al. 2004). While several counterparts have
been reported from ground based observations, at least 2 of them ``break up''
in deep HST images (Goad et al. 2002, Cropper et al.  2004) or are extended
(Immler et al. 2003) indicating that much of the optical flux comes from 
several
objects or is a star cluster. This variety can be seen in NGC4559, as is seen
clearly in the HST images (Cropper et al. 2004) the southern source (X-7) has 5
visible counterparts inside the Chandra error circle, while in the field of
the northern ULX (X-10) there are no counterparts. At present there are several
other ULX with possible optical counterparts,  2 with a probable O-star
counterpart (Liu et al. 2002b, Goad et al. 2003), one with a globular cluster
in a spiral galaxy (Wu et al. 2002)  and several with globular clusters in
elliptical galaxies (Angelini et al. 2001, Maccarone et al. 2003).

However the sensitivity of HST is such that only the most luminous stars can be
recognized at $D<15$ Mpc. There has not yet been  statistical work on the
likelihood that the  counterparts are real.  While the random association of a
luminous O star with a ULX is unlikely, the fact that ULX are often near star
forming regions raises the random probability considerably.  As opposed to the
early searches for galactic and active galaxy counterparts to X-ray sources,
none of the ULX counterparts  show ``unusual'' optical colors or variability.
This considerably increases the difficulty of directly associating the  optical
counterpart with the ULX.

 The optical studies have clearly shown that  the ULXs have very high X-ray to
optical ratios. X-ray selected AGN from the Rosat all sky survey tend to have
log $F(x)/F$(opt)$\sim$1, which for the brighter ULXs with $F(x)\sim10^{-12}$ ergs/cm$^2$/sec in
the 0.5--10 keV band would have optical counterparts at the 16mag level or 6--9
magnitudes brighter than usually observed. Thus the ULXs do not have the
optical properties expected if they were simple extensions of  AGN.

However low mass X-ray binaries in the Milky Way have $F(x)/F$(opt) $\sim$
100--$10^4$
consistent with the observed values in the ULXs. When these objects are
luminous most of the observed light comes from the accretion disk. If the
detected ULXs companions are high mass stars, the observed optical light is
consistent with that expected from a massive
 companion star with little or no contribution from an accretion disk. The
present data place only weak constraints on the optical emission from an
accretion disk  scaling from X-ray binaries in Milky Way.

\section{Radio observations of ULXs}
In the search for counterparts in other wavelength ranges it is natural to look
in the radio band. The radio data have excellent  angular resolution and
accuracy for help in finding an optical counterpart and contain unique
diagnostics for the  nature of the source (AGN, SNR, beaming, HII region, 
etc.).
Also the high sensitivity of the VLA allows the detection of intrinsically low
luminosity emission.  We  (S. Neff, N. Miller, RM) have  cross correlated  the
FIRST and NVSS radio catalogs with  Chandra/XMM for nearby galaxies.  This
allows a quick search for possible radio counterparts for further detailed VLA
analysis. So far we have $>12$ ``hits'' (with an angular distance
$\delta\theta
<1.5''$)
between FIRST radio sources and non-nuclear X-ray sources (we also have quite a
few with NVSS and XMM data with larger $\delta\theta$) and  several have ``good'' VLA
data.

 Because of the requirement that all sources be either in the FIRST or NVSS
surveys  their fluxes are greater than 1--2 mJy, or somewhat larger if the host
galaxy has bright diffuse galactic radio emission. A quick analysis shows that,
the ULX have a radio/X-ray ratio less than for Bl Lacs in general (Rector et al.
2002, Laurent-Mulhlensin et al. 1998) and that the radio/optical ratio is
larger; a typical Bl Lac optical to radio spectral index of 0.5  for a source
at a radio flux of 1 mJy  would predict an optical flux of $\sim$22--23 mag,
which is brighter than or equal to most of the sources. The observed
optical-X-ray ratios of the ULX are not  consistent with that of most Bl Lac
objects; that is the ULXs do not fall into the classification regions for Bl
Lacs in the X-ray-optical-radio plane. This  lessens the likelihood that the
ULX resemble these beamed objects.

A major surprise from our result is that a significant fraction of  the radio
counterparts to the ULX sources are resolved by the VLA. We had anticipated
that the radio emission would be point-like based on the original discovery by
Kaaret et al. (2003) of a radio counterpart to the ULX in NGC5408.  Kaaret et al.
claim  that the radio counterpart to the ULX is compact and thus would be
consistent with the radio emission  being due to relativistic beaming in the
ULX. The relative lack of point-like radio counterparts in our mini-survey is a
strong argument against relativistic beaming models.
In Table I we show the radio properties of several of the ULX sources. 
    Columns 3, 4 and 5 are the luminosity of the radio structures in units
of Cas-A (the most luminous SNR in the Milkyway), the radio power
in watts and the radio spectral index. 
\begin{table}[b]
\caption{ULXs.}
\label{table:1}
\begin{center}
\setlength{\tabcolsep}{1mm}
 \begin{tabular}{lcccccc}
\hline
NAME & $\begin{array}{c} {\rm Distance}\\{\rm   (Mpc)}\end{array}$&	Size (pc) & \# Cas-A's&	Radio Power &
$\begin{array}{l}{\rm Spectral} \\{\rm index}\end{array}$& Notes\\ \hline
$\hspace{-2mm}\begin{array}{l}{\rm NGC2782} \\ {\rm Starburst}\\ {\rm galaxy}\end{array}$ &  34	&80	&$\begin{array}{c}2800\\ 1100 \\ 600	\end{array}$&
$\begin{array}{c}{\rm 1.5E21}\\{\rm 6E20}\\{\rm 3E20}\end{array}$	
&$\sim -0.3$&	$\begin{array}{c}{\rm 3\ peaks}\\
{\rm Resolved\ \ arc }\\{\rm of\ \ emission}\end{array}$\\\hline
NGC3877&	12	&$260\times170$&85	
&4e19&$-0.1$&$\begin{array}{c}\sim 7'' \ {\rm from} \\{\rm nucleus{\mbox -}jet?}\end{array}$\\\hline
NGC4314	&13&	$<125$	&20 (each)&	1E19 (2)&	$-0.4$&	2 parts\\\hline
NGC4449&	2.8&	$8\times4$&	$\sim 10$&	5E18&	Steep,1.7&	
SNR\\ \hline
NGC4490&	6.6&$\begin{array}{c}	13 \ {\rm (core)}\\
\sim 65 \end{array}$&	$\sim6$&	3E18&	$-0.5$&$\begin{array}{c}	
{\rm Core }\\{\rm halo/double}\end{array}$\\ \hline
HoII	&2&	$40\times30$&	$\sim 1$&	5E17&	$-0.3$&\\ \hline
NGC3256	&	&$90\times 270$&	$\begin{array}{c}1000\\100	
\end{array}$&&	$-0.8$ (ULX)&$\begin{array}{l}	{\rm Twin\, AGN} +\\ {\rm
ULX}\end{array}$\\ \hline
NGC5408&	4.8	&&	1.5	&&	$>-1$	&Kaaret et al.\\ \hline
\end{tabular}
\end{center}
\end{table}

A simple calculation shows that the sensitivity of FIRST limits all the radio
counterparts, at distances greater than  3 Mpc, to be more than 3 times as
luminous as the most luminous radio source in the Milky Way, the supernova
remnant Cas-A, and that to be resolved by the VLA the objects have to be rather
large ($>20$ pc at 3 Mpc distance). Thus their mere detection and resolution
show that these objects are very luminous and very large for SNR or HII
regions. Our results have been confirmed by Wang et al. 2003 who find several
large and luminous resolved radio counterparts in NGC 3556, which because it is
not a FIRST source, did not enter our sample.

The morphologies of the radio sources vary (Fig.~1) but some look distinctly
like double radio sources.  Their  maximal cooling times (if emission is
thermal like in HII regions) is $<3\times10^8$ yrs if there is no continuous energy
injection and an order of magnitude shorter if they are synchrotron sources in
equipartion. So far only one source has clear nature, the SNR in   NGC4449
(which at $D=3$ Mpc has a radio luminosity of 10$\times$Cas-A). While young SNR can be
as luminous as $L(x)>10^3$ Cas-A, this high a luminosity in an ``old'' SNR 
is highly
unexpected.

Some examples of the radio emission are\\
 NGC4631 : 

We see the general association of radio emission  and ULXs. The most
luminous radio/X-ray source has log $L(x)\sim39.7$ (.3--10 keV) and can be well fit
by a disk black body  model  with $kT$ diskbb =1.2 keV, and column density
$N(H)=2.6\times10^{22}$. This object is located  in the same place as  the CO wind
discovered by Rand (1999). Rand argues that the presence of the CO wind
required roughly  $10^{54}$ ergs of kinetic energy, considerably more than can be
provided by a single supernova.  The brightest X-ray source in the galaxy  (to
the west---not in the image in Fig.~1) has  $F(x)= 2.6\times10^{-12}$ and  a
 $L_{\rm bol}
\sim 3.8\times10^{40}$ ergs/sec and is well fit by a simple power law with column density
 $N(H)=3.4\times 10^{21}$.\\
Holmberg II (UGC4305): 

This dwarf galaxy contains a very luminous source with
$L_{\rm bol}\gtrsim 2\times 10^{40}$ ergs/sec (Dewangan et al. 2004). The VLA source is coincident with
the Chandra source  ($\pm0.5''$) and  is resolved, with an effective size of
$2\times 1.4''$ at 4.86 Ghz and smaller at 1.4 Ghz (for an equivalent size of 
$\sim20\times25$ pc).
Its flat radio spectral index $-0.29\pm0.35$, is unusual for a thermal supernova
remnant and the NVSS  flux of 15mJy corresponds to a radio luminosity of
12$\times$Cas-A, while the  VLA resolved flux is $\sim 0.1$ of this, consistent with the
observed morphology.
\begin{figure}[t]
\parbox{\halftext}{
\centerline{\includegraphics[width=6.6cm]{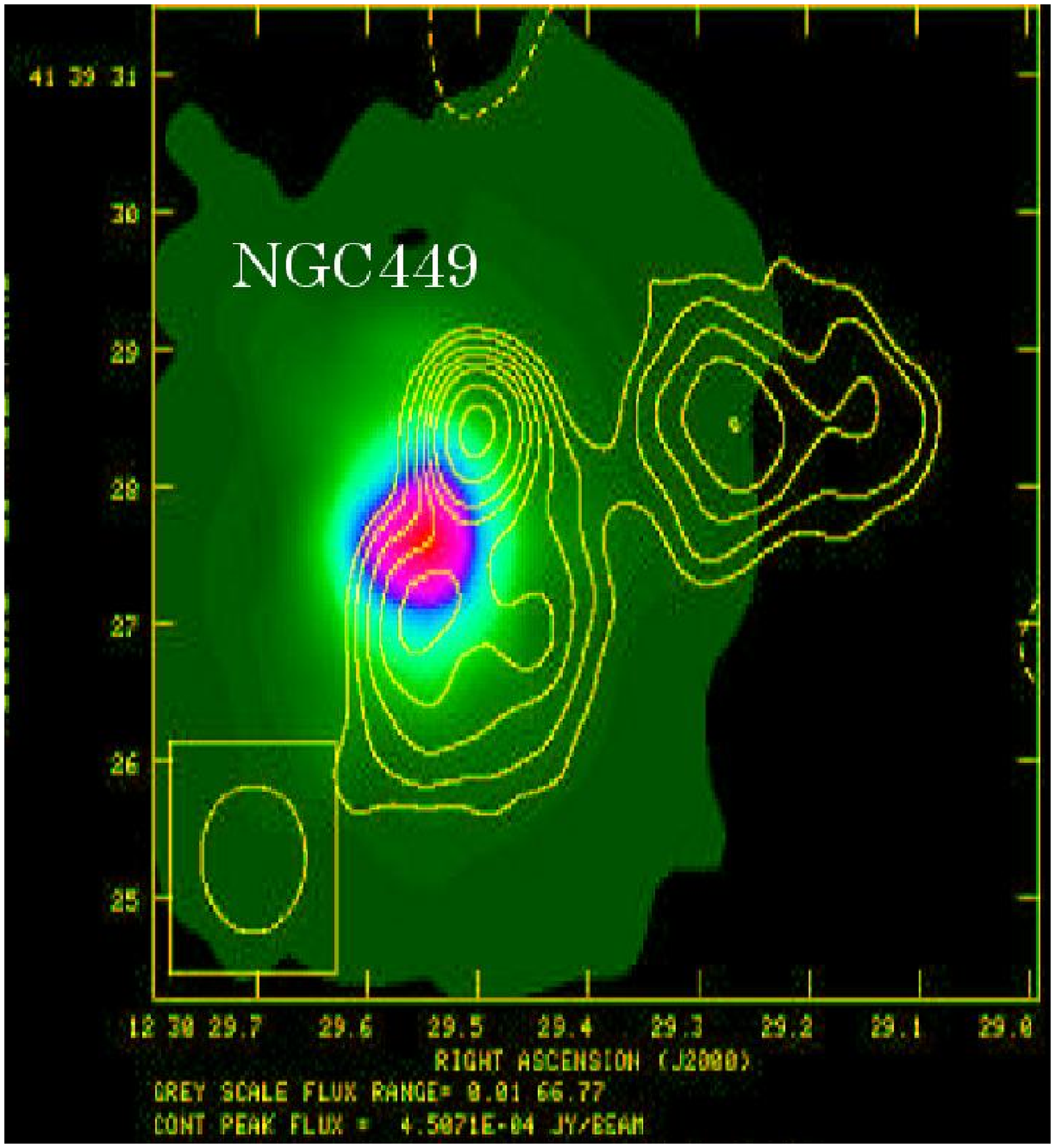}}
\vspace*{1cm}
\centerline{\includegraphics[width=6.6cm]{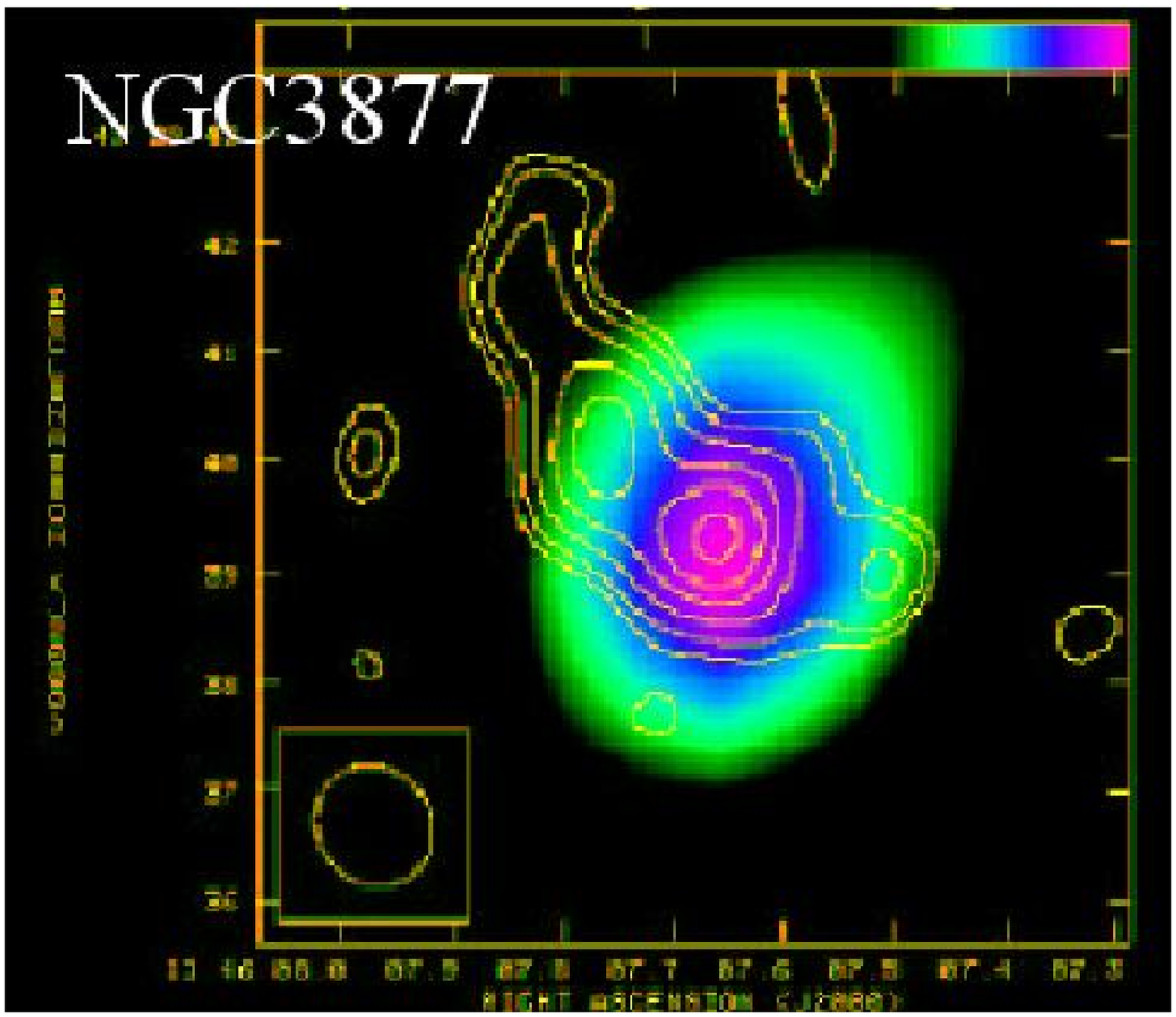}}}
\hfill
\parbox{\halftext}{
\centerline{\includegraphics[width=6.6cm]{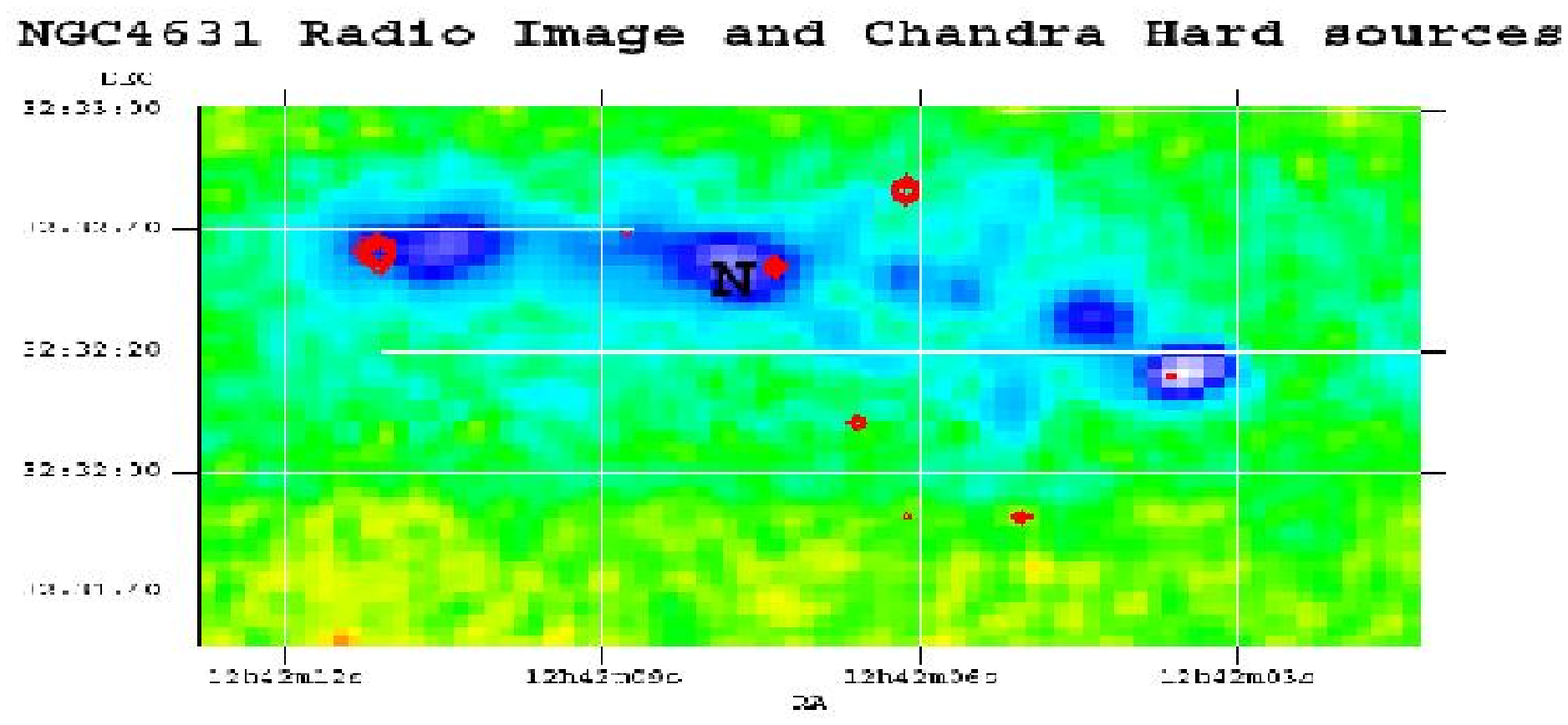}}
\vspace*{1cm}
\centerline{\includegraphics[width=6.6cm]{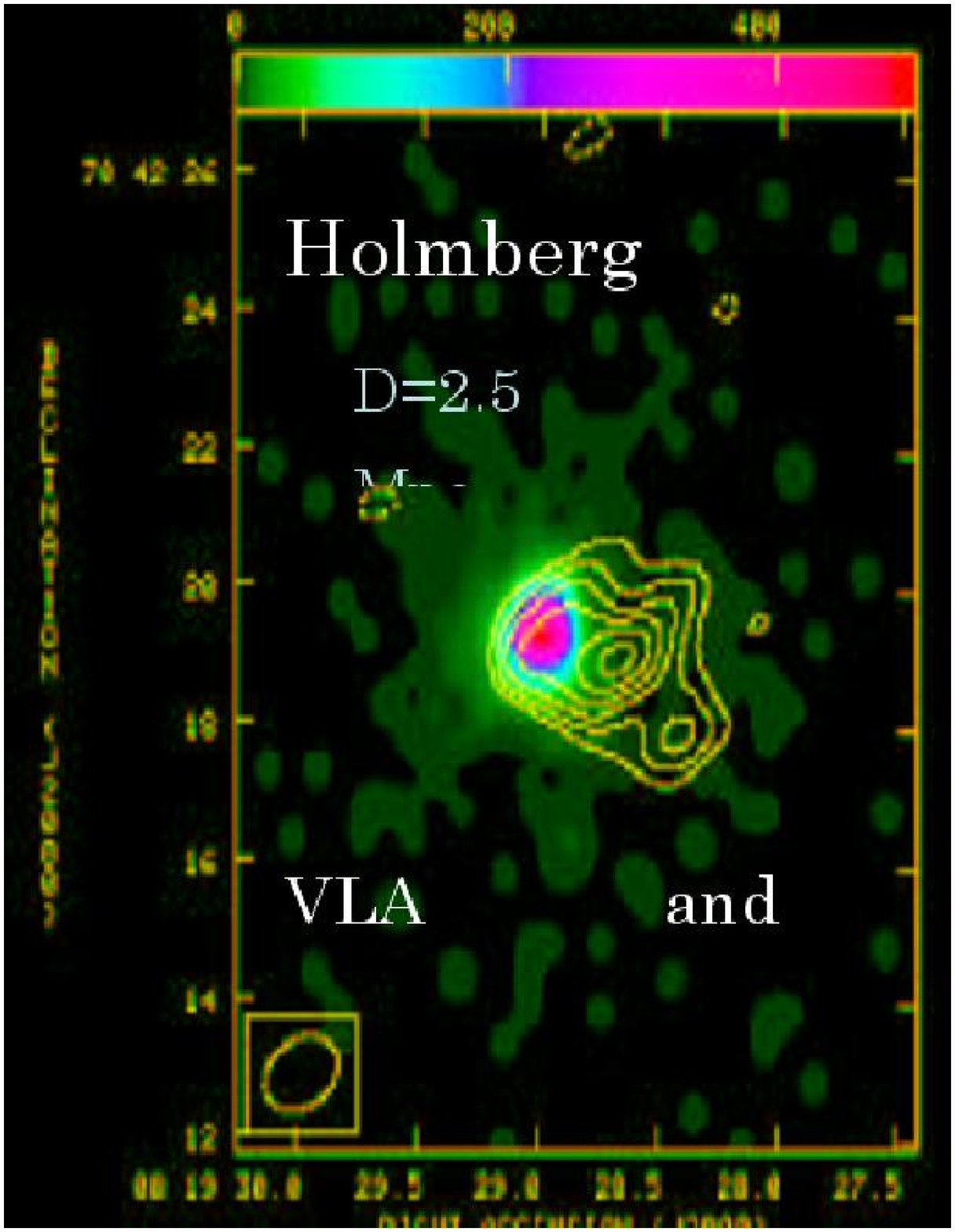}}
}
\vspace*{0.5cm}
\caption{
The VLA radio images of 4 ULXs. The radio data are shown as contours
superimposed on the grey scale Chandra image.}
\label{fig1}
\end{figure}

The XMM spectrum of the ULX shows a strong soft component (cf. Dewangan et al.
2004) which can be well fit by a diskbb model. The presence of a strong black
body component, for this luminous ULX, inside a bright extended radio source is
quite inconsistent with beaming in our line of sight!  One should also note
that the X-ray spectrum of the radio emitting ULX in NGC5408 also requires a
similar black body component (see below).\\
NGC4314:

This object has one of the most spectacular examples of a ring like
optical emission line structure as seen by HST and a similar shaped radio
structure surrounding the nucleus. The observed radio luminosity is  too large
in the  ``knots'' to be simply the  sum of a ``reasonable'' number of SN . The
sources associated with the radio emission are X-ray sources X-1  and X-3 and have X-ray
luminosities of  $L(x) \sim 3\times10^{39}$ ergs/sec and $L(x)\sim 7\times10^{38}$ ergs/sec
respectively.\\
NGC3877:
 
For this galaxy at $D=17$ Mpc ($1''=$ 83 pc) the  VLA source is exactly coincident
with the Chandra source ($\pm 0.5''$). The radio source has a flux of $\sim$3mJy and is
 resolved with a rough size of $2\times .4''$ at 4.86 Ghz and smaller at 1.4 
Ghz with a
flat spectral index of $ -0.13\pm0.35$. This translates to  a size of
150$\times$300 pc
and a luminosity of 80$\times $Cas-A.  The object is only $\sim 7''$ away from optical
nucleus and has 5 Chandra observations with an average luminosity of $L(x)
\sim6\times10^{38}$. There is  nothing obvious in the HST images indicating that the
radio emission is not due to a giant HII region.\\
NGC4490:

For this galaxy at $D=8$ Mpc $1''=$39 pc and the   VLA radio image  is
coincident with the Chandra source ($\pm0.5''$). The source  is resolved 
with a size of $2\times .4''$
at 4.86 Ghz about 75$\times$150 pc and the spectral index is flat $-0.13\pm 0.35$. With a
flux of $\sim$3mJy it is  15$\times$Cas-A.  The X-ray flux varies between Chandra and XMM
epochs with the Chandra $L(x) \sim 8\times 10^{38}$.

\section{Nature of host galaxy}
ULXs can occur in any galaxy but are most frequent in rapidly star forming
galaxies.   However they can also occur in dwarfs, such as Holmberg II, and in
elliptical galaxies with little present day star formation and in regions far
from any obvious star formation (e.g. NGC1313 X-2 and NGC4559 X-10). In
elliptical galaxies it seems as if  the maximal luminosity is $10^{40}$ ergs/sec,
below the maximal luminosities seen in actively star forming galaxies. In
NGC720, a nearby giant elliptical with no star formation, the number of ULXs is
comparable to that of active star forming galaxies (Jeltema et al. 2003) and at
least one is in a globular cluster. ULXs in globular clusters in elliptical
galaxies are the most challenging to the association with star formation since
these are old systems with no star formation in several gigayears.

In NGC4649, a nearby giant elliptical with no recent star formation ULX 69
(Colbert and Ptak 2002) has a good XMM spectrum. It is well fit by a power law
(a diskbb is ruled out for the hard component) with an indication of  a  black
body component and a bolometric luminosity of $\sim 10^{40}$ ergs/sec.  If the black
body is physical it implies a size of $6\times10^3$ km which gives  mass of
$\sim 1000M$ if
the object is in the galaxy. There is a counterpart on the XMM OM UV data and
the POSS plates and this possible ULX is either in a globular cluster or,
alternatively, it is a background AGN only $3'$ from a galaxy center. This
raises the possibility that some of the ULX are projections of background AGN
against the host galaxy. Based on the observed ratio of X-ray to optical fluxes
for AGN this is only likely for the optically brighter ULX counterparts, as has
been recently noted to one of the sources near NGC720 (Arp et al. 2004).

\section{Time variability}
Time variability is frequently observed in the ULXs, arguing that most of them
are single compact objects, rather than a sum of numerous lower luminosity
objects in the same object.  Because in the pre-Chandra/XMM era the count rates
were usually low, most of the time variability data is on timescales between
observations  ranging  from months to years (Roberts et al. 2002).

Some of the ULXs may show X-ray periods, with data being published for   IC
342;  31 or 41 hrs (Sugiho et al. 2001), the Circinus galaxy ULX with 7.5 hrs
(Weisskopf et al. 2003) and M51 X-1; with a 2.1 hr period (Liu et al. 2002). The
periods in the Circinus galaxy and M51 objects appear to be due to eclipses,
indicating that the companion is a low mass star.

\begin{figure}[t]
\parbox{\halftext}{
\centerline{\includegraphics[width=6.6cm]{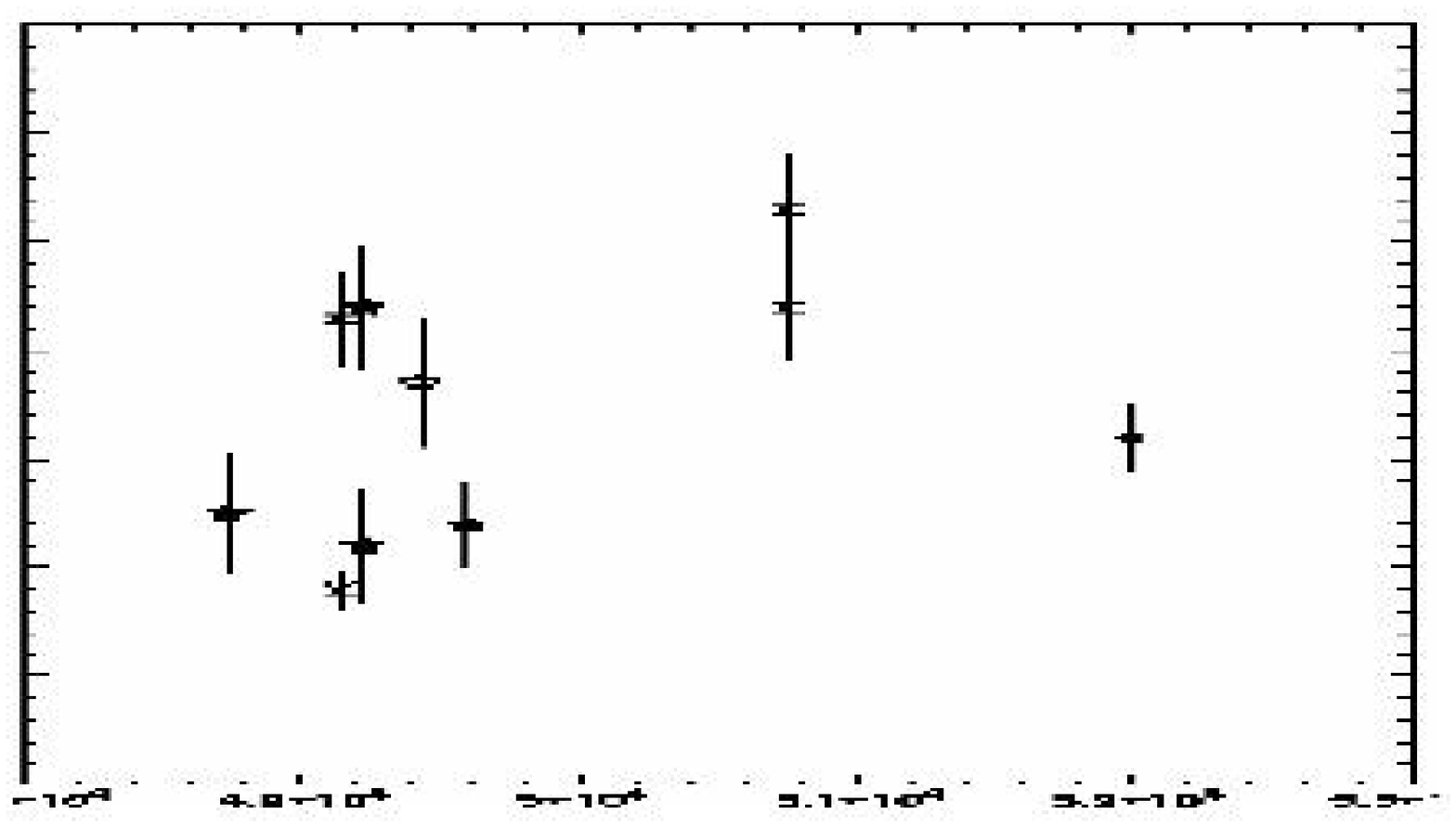}}
\vspace*{1cm}
\centerline{\includegraphics[width=6.6cm]{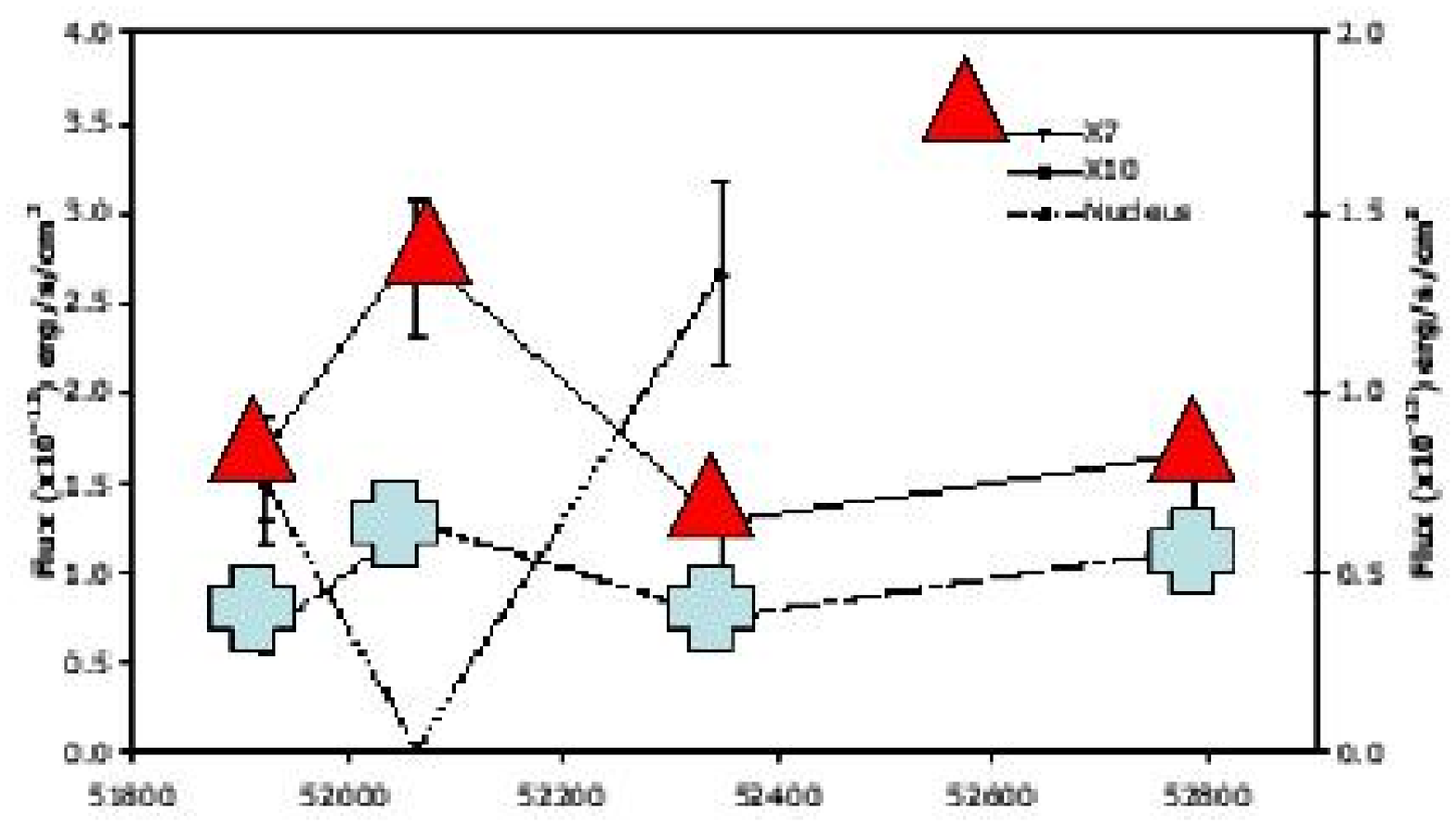}}}
\hfill
\parbox{\halftext}{
\centerline{\includegraphics[width=6.6cm]{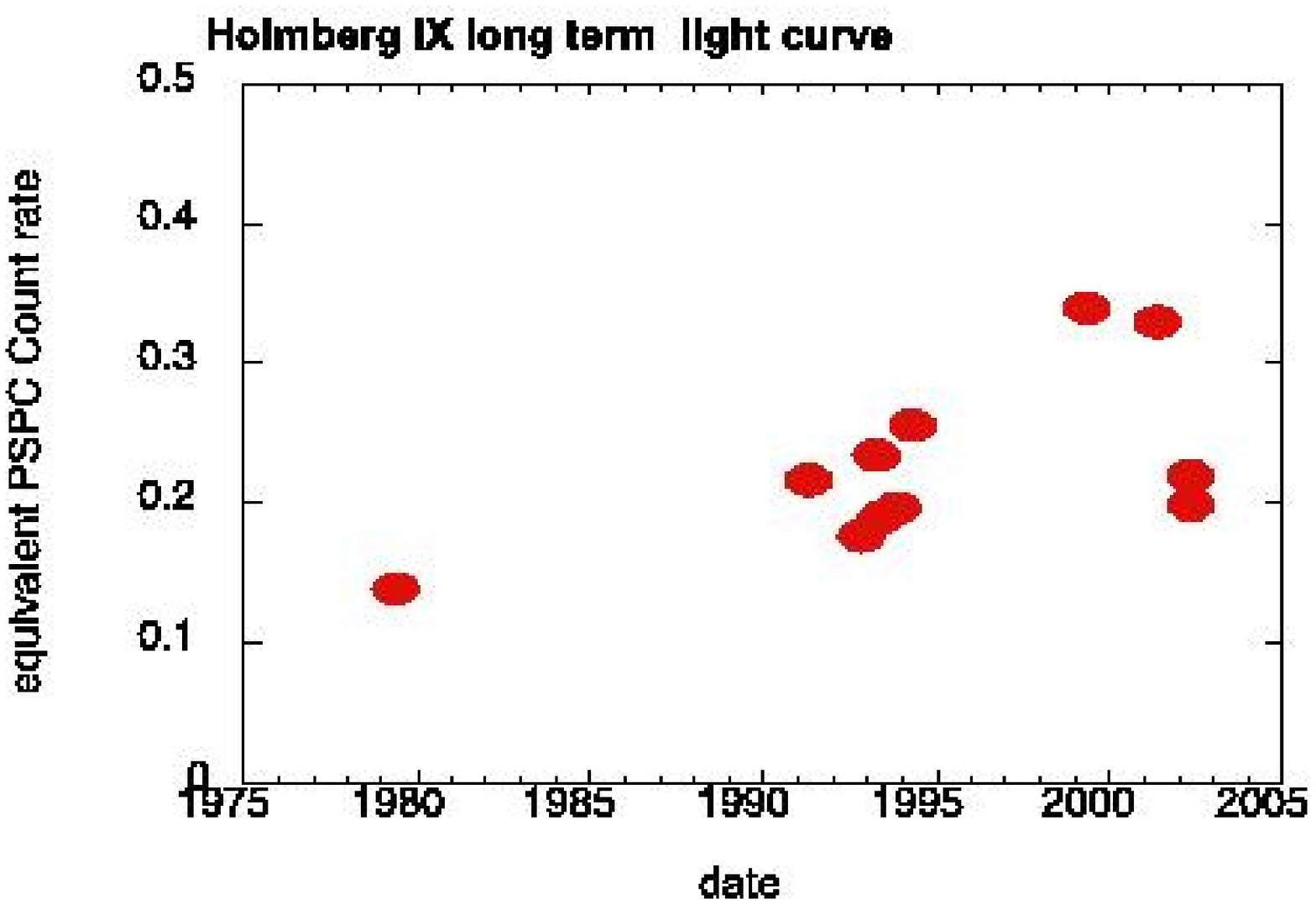}}
\vspace*{1cm}
\centerline{\includegraphics[width=6.6cm]{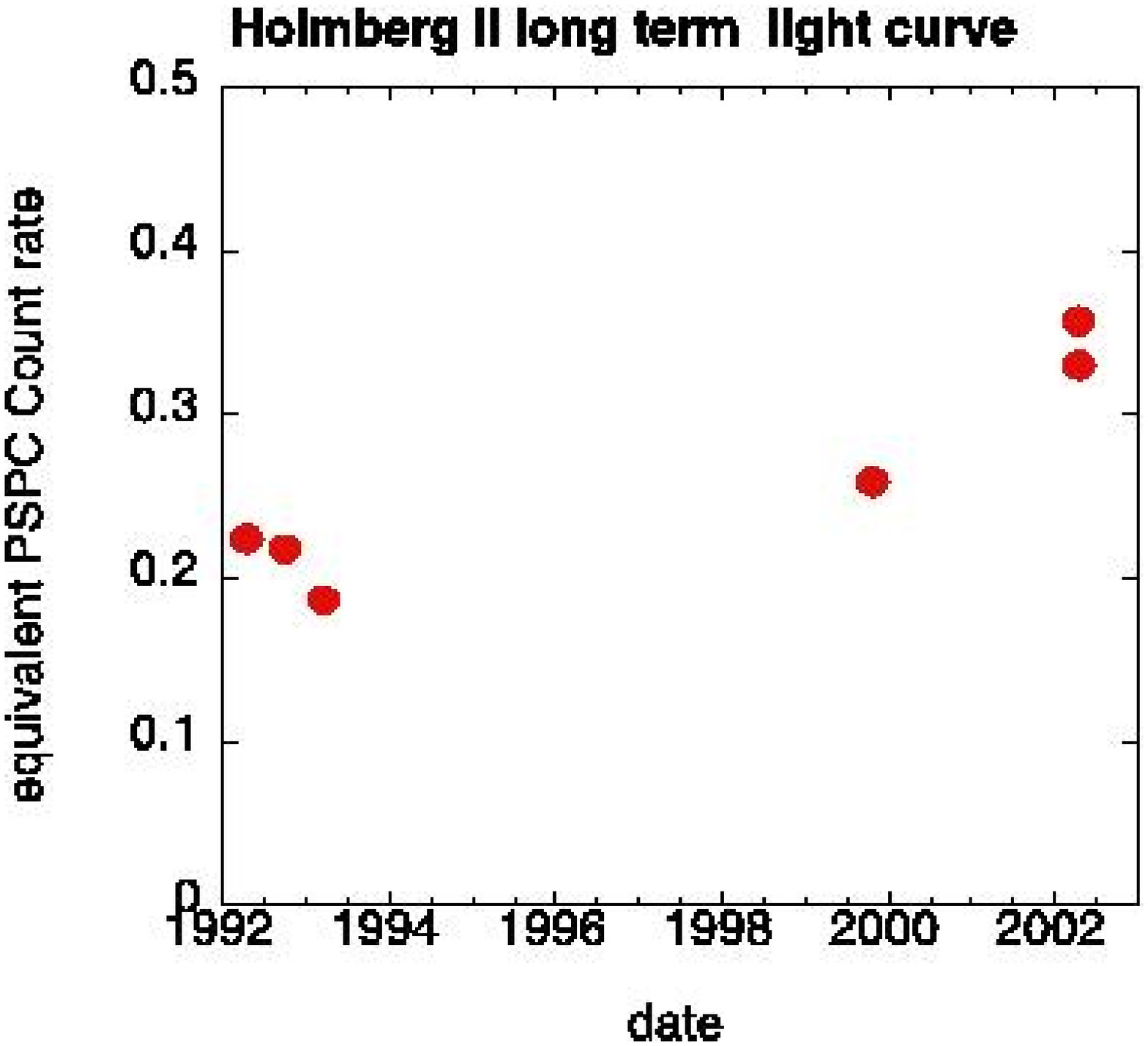}}}
\vspace*{0.5cm}
\caption{
The long time scale variability of 5 ULX; M81 X-9, Holmberg II,
NGC2276 and the two ULX in NGC 4559.}
\label{fig2}
\end{figure}

\begin{figure}[t]
\centerline{\includegraphics[width=8cm]{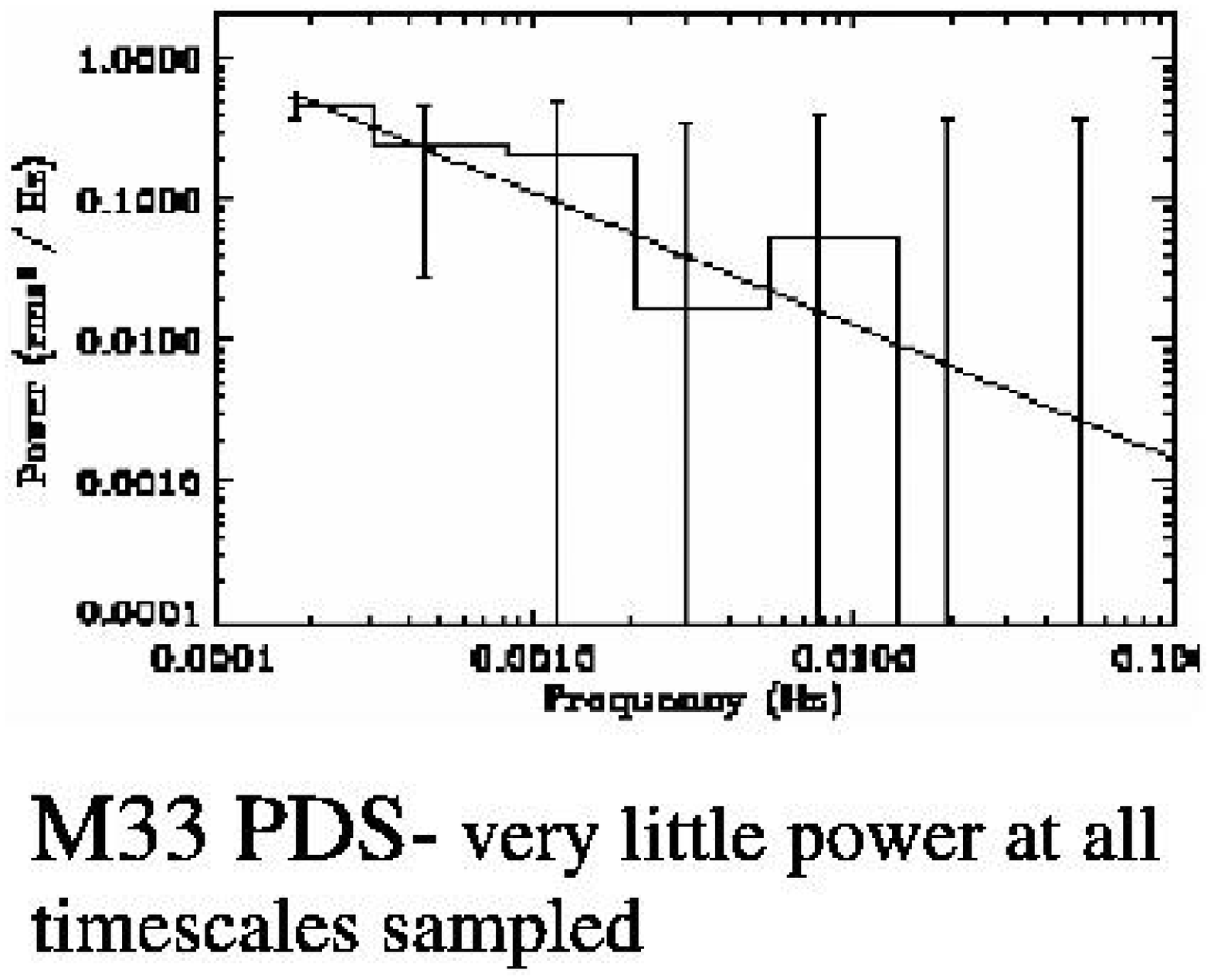}}
\centerline{\includegraphics[width=8cm]{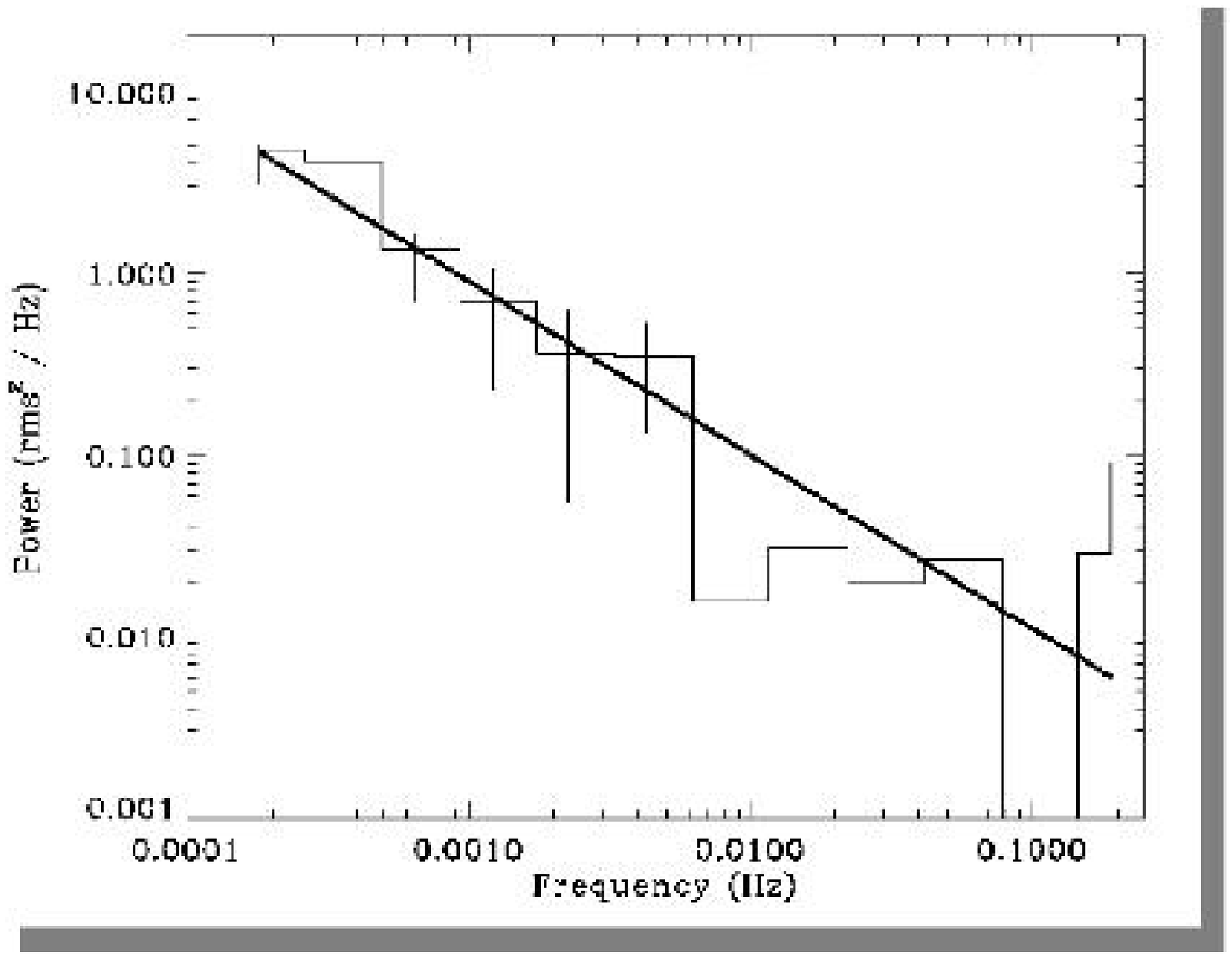}}
\caption{
The PDS for the central source in M33 and the Circinus ULX
(Strohmayer p.c.) showing the low overall power in these objects and the
absence of  QPOs and a break in the PDS.}
\label{fig3}
\end{figure}

While most ULXs vary, many show low amplitude variability on long time scales
which is very different than galactic  black holes or Seyfert galaxies (except
for the black hole candidate LMC X-1 !) (Fig.~2). It is not clear if the
apparent low level of variability is due to the sporadic nature of the
observations or is a true property of the objects --- this can only be answered by
denser sampling of the time series. There are also quite a few examples of
objects with large amplitude variability (e.g Strickland et al. 2001) including
a few ``transients''.

Many galactic black holes exhibit ``quasi-periodic oscillations'' (QPOs) which
are clearly associated with the accretion disk and represent characteristic
length scales close to the black hole.   If  the QPO frequency is associated
with the Kepler frequency at the innermost circular orbit for a Schwarzschild
black hole, then the frequency of  the observed QPO in M82 (Strohmayer and
Mushotzky 2003) of 0.06 Hz translates to an upper limit on the mass of $1.9\times10^4M_\odot$, consistent with observed luminosity and an efficiency of $\sim0.1$, similar to
that of most AGN and galactic black holes.

It has long been known that galactic black holes exhibit characteristic power
density spectra (PDS), flat at low frequencies and steep at high frequencies
and that, in some sense, this is a defining character of these objects. The PDS
for AGN shows a similar form, with the break frequency scaling as  the mass of
the object over a factor of $10^7$ in mass  (Markowitz et al. 2003). Only XMM has
the signal to noise to measure an accurate PDS for  $\sim$ 5--10 ULXs if their PDS
scales from Cyg X-1 or Seyfert galaxies. The PDS for several XMM sources are
well sampled, with good signal to noise.  Preliminary analysis for several ULXs
shows that many of them have  low overall power (Fig.~3) with no more QPOs (yet)
and thus the ULX, in general,  do not have the ``characteristic'' BH power
spectra. However there is a singular exception, the ULX X-7 in NGC4559 (Cropper
et al. 2004). X-7 has a ``classical'' Cyg X-1 power spectrum, flat at low
frequencies and steep at high  and with an RMS variability of 37\% it is very
similar to Cyg X-1.  However its  break frequency of $\sim  28$mHz is 50 times
longer. Scaling  that break frequency to mass (as for AGN and Cyg X-1) gives
$M\sim 10^3M_\odot$. In the same galaxy the other ULX, X-10, has a   steep power law PDS
with  little high frequency power and no characteristic frequency.

We believe that the search for characteristic frequencies is, at present, the
most productive way of determining the nature of the ULX. However to utilize
this technique requires moderately long exposures of sources with $>1$ ct/sec.
Since all black holes seem to show the same characteristic PDS a search for the
characteristic frequency is crucial for a large sample. If it is indeed true that
the form of the PDS is universal, then the absence of a low frequency break in
several of the measured PDS implies rather large masses.

\section{X-ray spectra}
To first order the X-ray spectra of accreting black holes can be modeled, for
most AGN and galactic black holes by two forms, a pure power law and a power
law with a disk black body component. If  the X-ray spectrum of the ULXs can
also  be well described by this model, then a simple relation exists  between
the color temperature and luminosity of the disk black body component  and  the
mass of the object is (Ebisawa et al. 2002)
$T_{\rm color} \sim 1.3$ keV (($T_{\rm color}/T_{\rm eff})/1.7)(M/M_{\rm
edd})^{1/4}(M/7M)^{-1/4}$ where $T_{\rm color}$ is the
effective diskbb temperature fitted to X-ray spectra. Since the  implied masses
of the ULXs are greater than  $20M$, one expects $T_{\rm color} < 1$ keV. However, based
on ASCA data (Colbert and Mushotzky 1999, Makishima et al. 2000)  many of the
ULX  have $T_{\rm color} > 2$ keV inconsistent with high masses.

Detailed calculations (Ebisawa et al., these proceedings) indicate that this ``color
temperature'' problem is not generally solved in a Kerr metric. Thus the ULXs
are too ``hot'' for their inferred ``Eddington limited'' mass and either the
spectral model, masses or interpretation is wrong. It is clear that the ASCA
data were not very ``wrong'' since high S/N XMM data confirm that some sources
have curving spectra. However the  spectral fits are not unique and these
spectra can be fitted either by the diskbb models or by Comptonization spectra,
as originally pointed out 15 years ago for LMXB spectra (White et al. 1988).
For example in the case of   M81 X-9, the XMM high signal  to noise data are
equally well fit by  power laws with black body spectrum,  Comptonization
models and diskbb + power law, while for Holmberg II the BMC models of Shrader
and Titarchuk, Compotonization and black body spectra and power law fit well
while the diskbb   with a black body soft component is a poorer fit.  If the
spectra  can be interpreted as being due to Comptonization, as for low mass
X-ray binaries, this alleviates the  problems with high $kT$ disk black body
models (see Kubuota, these proceedings). The very high signal to noise M33 X-1
spectrum (Fig.~4) can be very well fit by a Comptonized spectral form with low
temperature and high optical depth. Only higher energy data can test which, if
any, of these models are the correct description.

\begin{figure}[t]
\centerline{\includegraphics[width=8cm]{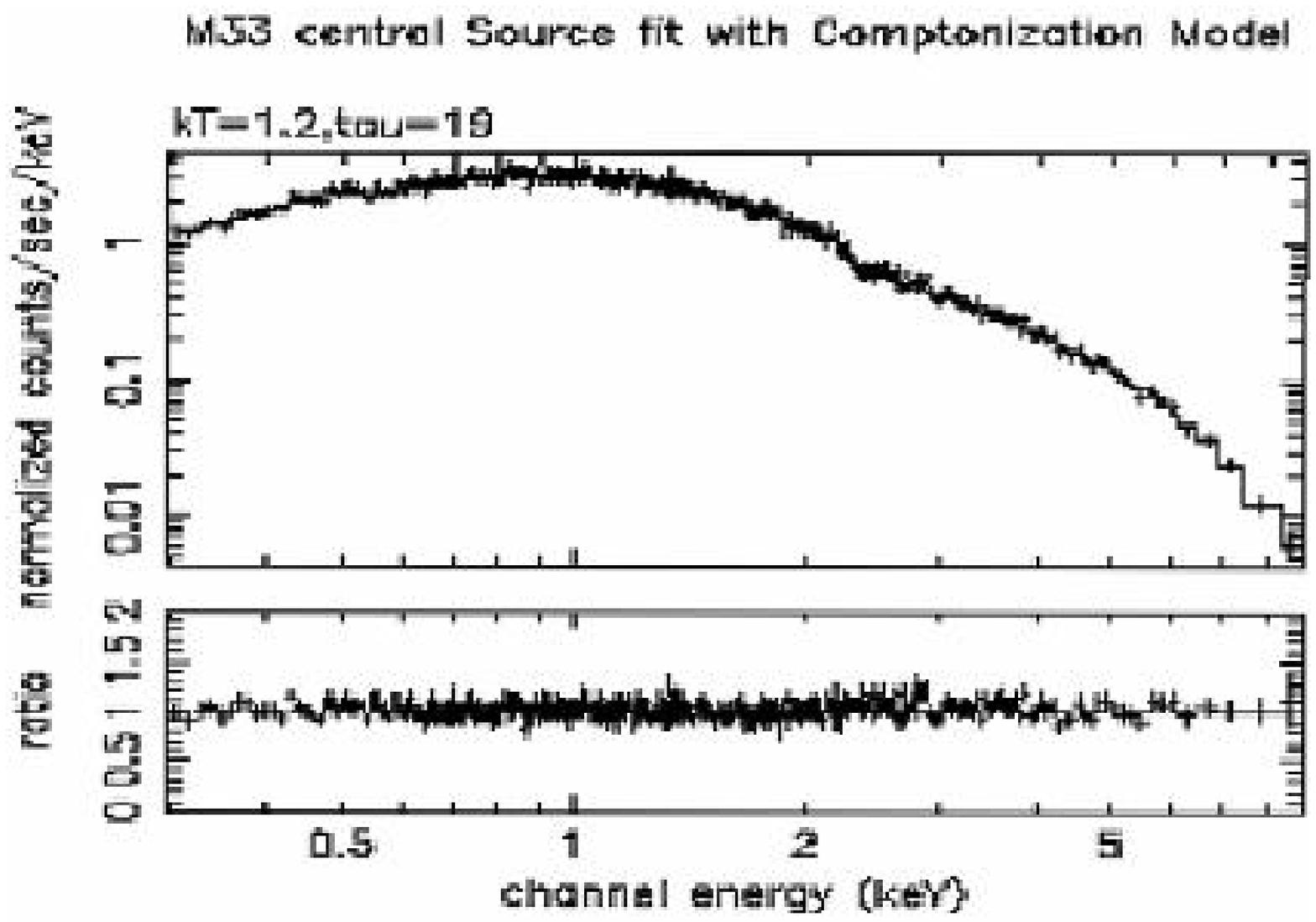}}
\vspace*{1cm}
\centerline{\includegraphics[width=8cm]{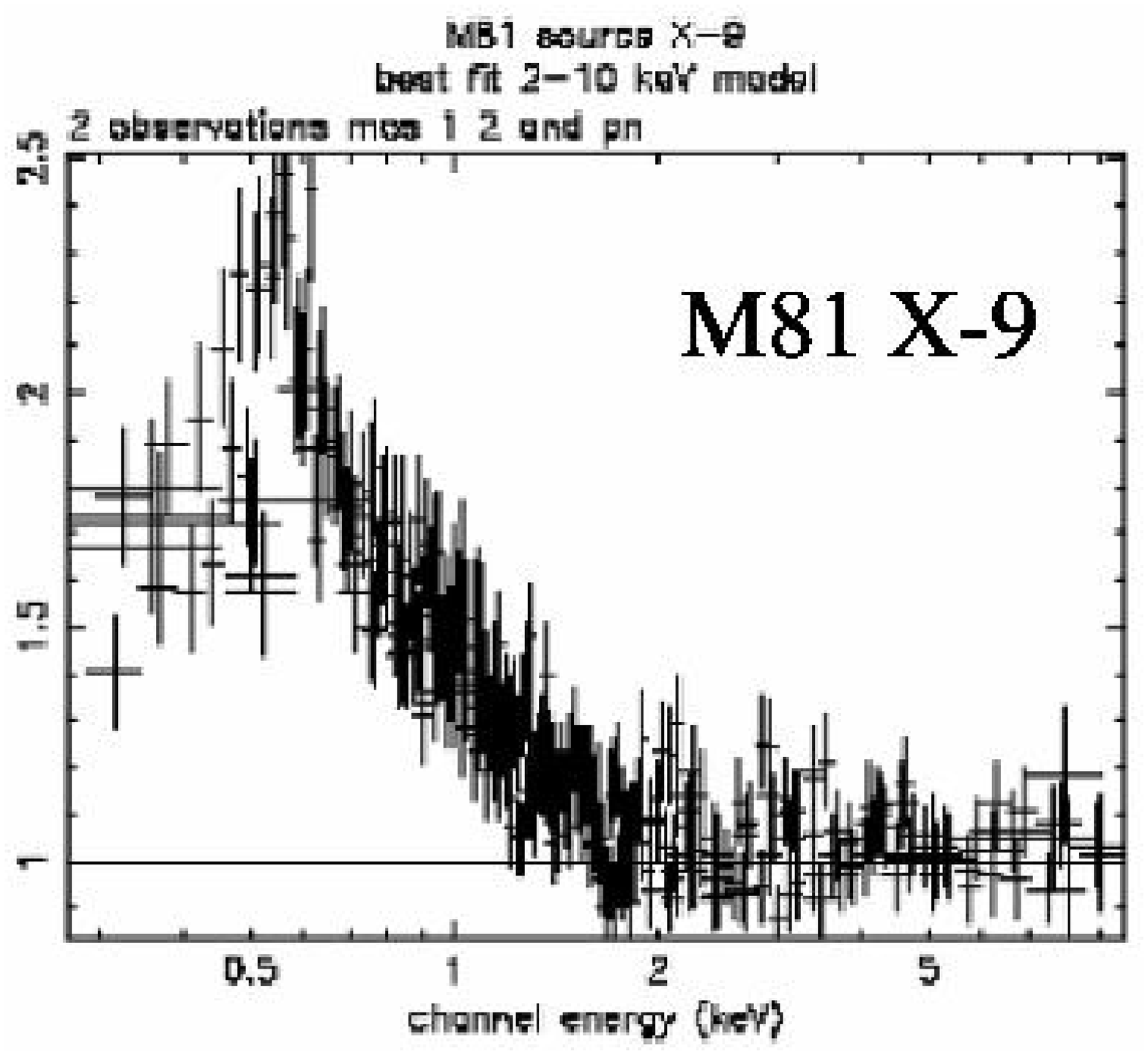}}
\caption{
Spectral fits to XMM data for several high signal to noise observations. 
(a) The fit to  M33 central source with the Sunyaev-Titarchuk 
comptonization models with $kT\sim 1.22$ and $\tau\sim 10$. While this is an excellent fit 
to the data, the fitted parameters are outside the formal range of allowed 
values for the parameters. (b) The ratio of the data to a power law for the 
ULX X-9 in M81, illustrating the need for a strong soft component. 
}\label{fig4}
\end{figure}

Given the uncertainties in the form of the continuum, factors of 2--3
uncertainty in the bolometric correction exist; in particular for fits with a
steep  power law the bolometric luminosity diverges at low energies. Since the
optical fluxes are so low we know that the power law form cannot continue to
arbitrarily low energies, contrary to the case in AGN.

With XMM quite a few sources require soft components which can be well fit by
black body spectra  with $0.1<kT<0.3$ keV. The low temperature of the soft
components is hard  to detect with ASCA and Chandra ACIS spectra because of
their relatively poor soft response. The exact value of $kT$ depends on the
model used for the hard component and the form of the low energy absorption. In
some cases, but not most of them,  the adoption of a low abundance absorber
removes the need for a soft component. However not all luminous sources require
soft component. It is interesting that the range of inferred black body
temperatures from these fits 0.1--0.2 keV is very similar to what is found for
narrow line Seyfert galaxies, despite their apparent $10^4$ difference in mass.

As an example of a detailed spectral fit, we consider the spectrum of X-7 in
NGC4559 (Cropper et al. 2004). The XMM spectra have $\sim 20,000$ counts and are well
fit by a  power law ($\Gamma =2.23$) and   ``black body'' like component of
$ kT\sim 0.14$
keV. The luminosity and temperature in the BB component give an effective size
of  $R \sim 3\times10^9$ cm,    which for the King and Pounds (2003) wind model gives a
model mass  $M\sim 2\times 10^3M_\odot$, while the radius of a disk black body model is
$R_{\rm diskbb}=1.2\times10^9$ cm; if this corresponds to $6R_G$ then the black hole mass is
$M\sim 1.6\times10^3 M_\odot$.  The BMC model (Titarchuk and Shrader 1999) gives a  similar mass.
Using these fits one derives a  bolometric luminosity of  $L_{\rm bol}\sim 
6\times 10^{40}$ ergs/sec
or  $L\sim 0.1 L_{\rm Edd}$ for the masses estimated from the spectral fits and the break in
the power density spectra.   There is no Fe K line, with an upper limit on the
$EW<100$ eV for a narrow line and 200 eV for a broad line.\\
Fe K lines:

The XMM data for bright sources with good $S/N$ typically do not show Fe lines,
with the sole exception of M82 and the Circinus dipper. For the best spectra
the upper limits are $\sim 50$ eV; for several of the brigher object the upper limits
are $<100$ eV for narrow lines. These limits are already interesting when compared
to similar quality ASCA data for many Seyfert I galaxies. There are strong
hints of oxygen lines in several sources but it is not clear if it is diffuse in
origin or related to the ULX itself (e.g. Dewanag et al. 2004). For M82 and
Circinus dipper the Fe K line is complex and  broad. The EWs are $>100$ eV; (in
Circinus 2 lines of $\sim $180 and 320 eV EW, in M82 the limits are $>130$ eV EW  for a
broad Gaussian and $\sim250$ eV for a diskline.

The existence of  strong and/or broad Fe K lines shows that, in these two
objects, the  continuum is not beamed since in order to produce lines of this
strength requires the  interaction of the X-ray continuum and a large solid
angle of  cold material, which is not available in the beaming scenarios.

\section{Conclusion}
There is no direct evidence for beaming in any of the ULX: that is, they do not
have the timing behavior, X-ray spectra or the broad band radio/X-ray-optical
spectra characteristic of known types of beamed sources. In 4 sources there is
direct  evidence against beaming (M82 has both a QPO and a broad Fe line, NGC
4559 X-7 has a Cyg X-1 PDS, 2 sources are eclipsing (M51 and Circinus) and 2
(M82 and Circinus) have broad Fe lines). It is also becoming clear that  many
sources have a luminous  soft black body like component. If this is due to an
optically thick accretion  disk (as is often assumed) it provides  indirect
evidence against beaming.

There is evidence for high intrinsic luminosity in several objects  from the
ionization of the associated optical nebulae and the presence of  black body
components. Some of the X-ray spectra show significant high energy curvature,
different from that of known AGN and almost all of the known galactic black
hole spectra, but similar to, but hotter, than that of LMXBs. Similar to LMXB
most of the ULX do not have strong Fe K lines.
Most of the X-ray PDS do not resemble that of Cyg X-1 or AGN with significantly
less power over the frequency ranges sampled and only one source shows a PDS
form of AGN or galactic black holes.  There are associated luminous, large
radio sources whose origin is not clear;  this represents  a new ``type'' of
object  associated with the ULXs. While their direct physical association has
not yet been established, we are unaware of any such radio sources not
associated with ULXs.

Thus the ULXs do not ``look like'' scaled up GBHCs or scaled down AGN nor like
beamed versions of either one. The sum of the results do not ``hang together''.
Either we are dealing with 3 or more ``new'' types of objects or we have to
re-think what a black hole should ``look like''. I suspect that what we are
detecting in many of these objects is Comptonized optically thick spectra, of
relatively low temperature  rather than the low optical depth, high temperature
spectra characteristic of most AGN and galactic black holes in the low state.
These are very exciting times for ULX research and I fully expect that there
will be remarkable changes in our understanding of these objects in the next
few years.

\section*{Acknowledgements}
I would like to thank my collaborators at GSFC, Todd Strohmayer, Neal Miller
and Susan Neff for major contributions to this work, and my NGC4559
collaborators, especially Mark Cropper and Roberto Soria for communication of
results prior to publication. I would also like to thank the organizers for a
most exciting meeting, which allowed many different ideas to be brought forth.
I would also like to thank Professors K. Koyama and S. Mineshige for a  beautiful
introduction to Kyoto.

\end{document}